# Kinematic Discriminants of Deceleration Behavior Modes in Car-Following: Evidence from NGSIM Trajectory Data

ENI SOLOMON LAUGHTER[1]*
[1] COLLEGE OF TRANSPORTATION ENGINEERING, CHANG'AN UNIVERSITY, XI'AN 710064, CHINA
*CORRESPONDING AUTHOR: ENISOLOMONLAUGHTER@GMAIL.COM

## ABSTRACT

Gap-closing rate and visual looming swap discriminative dominance depending on deceleration intensity — a finding that reconciles a long-standing conflict in the car-following literature and challenges the spacing-centered assumptions of traditional driver behavior models. This study presents a two-stage analytical framework that distinguishes between information availability (kinematic variables measurable in the environment) and information utilization (variables that demonstrably separate driver behavioral patterns), applied to 1,060,119 valid car-following observations from the NGSIM trajectory dataset (2,932 vehicles). Six kinematic features are extracted, deceleration events are detected under two threshold conditions ($-0.5$ m/s² and $-0.3$ m/s²), K-means clustering identifies behavioral modes, and one-way ANOVA with eta-squared effect sizes ranks each feature's discriminative power. Three key findings emerge: (1) threshold selection fundamentally shapes behavioral inference — the stricter threshold yields three interpretable modes while the permissive threshold collapses these to two; (2) hard braking prioritizes gap-closing rate ($\eta^2 = 0.715$) while moderate braking emphasizes visual looming ($\eta^2 = 0.574$); and (3) spacing headway is negligible ($\eta^2 \leq 0.014$) across both thresholds. These findings provide empirically grounded candidates for perceptual cue prioritization and carry direct implications for ADAS warning system design and autonomous vehicle control formulations.

## 1. INTRODUCTION

Traditional car-following models — including the Intelligent Driver Model (IDM), Full Velocity Difference Model (FVDM), and Gipps' model — assume that drivers regulate longitudinal behavior primarily through absolute spacing and velocity difference, treating these as the central state variables governing deceleration decisions (Cao et al., 2020; Jiao et al., 2020). Yet perceptual studies reveal a persistent conflict: (Mulder et al., 2005) found that drivers rely mainly on relative velocity for longitudinal control, whereas (Xue et al., 2022) and (Durrani et al., 2021) identified visual looming — the progressive optical expansion of a lead vehicle on the driver's retina as the gap closes, quantified as inverse TTC ($\tau^{-1}$) — as the dominant cue in collision avoidance scenarios. This contradiction has not been resolved within a single empirical framework, in part because both camps study different urgency contexts without a systematic method for comparing them. Compounding this, (Wang et al., 2022) demonstrated that structurally identifiable car-following models can produce identical trajectory outputs under different parameter sets, meaning trajectory data alone cannot determine which variables drivers actually respond to. The empirical basis for cue selection in existing models thus remains largely assumed rather than validated.

The present study addresses this gap by operationalizing a distinction that is theoretically recognized but rarely tested empirically: the difference between information availability — kinematic variables that exist and are measurable in the driving environment — and information utilization — variables whose variation demonstrably separates driver behavioral patterns. This distinction matters because drivers cannot directly observe computed metrics such as TTC or required deceleration; they respond to optical and kinematic patterns accessible through perception. By extracting six kinematic features from NGSIM trajectory data and determining which best discriminates between data-driven behavioral clusters through ANOVA effect size analysis, we identify kinematic discriminants of deceleration behavior modes. Here, a kinematic discriminant refers to a measurable variable of vehicle motion — such as gap-closing rate or optical expansion rate — whose variation across deceleration events best accounts for differences in behavioral cluster membership; it is kinematic in that it derives entirely from observable vehicle trajectory, and discriminant in that its explanatory power is quantified through the proportion of behavioral variance it explains, not through model-fitting assumptions about driver cognition.

This study does not claim to measure driver perception directly; rather, it provides empirically grounded candidates for perceptual cue prioritization, grounded in observed behavioral heterogeneity rather than model-fitting convenience. Six kinematic features are extracted from 1,060,119 valid car-following observations, deceleration events are detected under two threshold conditions — a primary threshold of $-0.5$ m/s², which targets deliberate hard braking episodes consistent with established definitions of significant deceleration responses (Feng et al., 2018; Hu et al., 2023), and a secondary threshold of $-0.3$ m/s², which captures moderate speed adjustments and enables systematic comparison of how threshold permissiveness affects behavioral inference. K-means clustering identifies distinct behavioral patterns, and ANOVA effect sizes rank which features explain behavioral heterogeneity. A systematic threshold comparison reveals a sample-size versus pattern-distinctiveness trade-off with direct methodological implications for trajectory-based deceleration research. The findings inform ADAS warning system design, rate-based autonomous vehicle control formulations, and more realistic driver behavior modeling.

The remainder of this paper is structured as follows. Section 2 reviews the literature across three themes: trajectory data as a foundation for behavioral inference, the contested empirical basis for perceptual cue prioritization, and prior behavioral clustering work alongside the research gaps this study addresses. Section 3 describes the methodology, covering data preprocessing, kinematic feature extraction and deceleration event detection, and the clustering and cue importance framework. Section 4 presents the results in two subsections — event detection and kinematic profiles, followed by behavioral clustering and ANOVA-based cue importance findings. Section 5 discusses the implications of the threshold sensitivity paradox, the cue dominance reversal, the negligible role of spacing headway, and the study's limitations. Section 6 concludes with the study's three primary contributions and directions for future research.

## 2. LITERATURE REVIEW

### 2.1 Trajectory Data as a Foundation for Behavioral Inference

High-resolution vehicle trajectory datasets — particularly NGSIM, sampled at 0.1-second intervals — have transformed the capacity for microscopic behavioral inference, enabling detection of traffic oscillations, shockwave propagation, and heterogeneous following behavior previously inaccessible through aggregated measurements (Li et al., 2020). (Chen et al., 2023) consolidated over 80,000 car-following events across five public datasets in the FollowNet benchmark, demonstrating that trajectory data support both physics-based and data-driven behavioral modeling across diverse traffic conditions. A critical methodological distinction exists in how researchers apply these data: some treat trajectories as inputs for calibrating pre-specified model variables (Cao et al., 2020; Jiao et al., 2020), while others adopt an inferential stance — using trajectory patterns to discover which kinematic variables drivers demonstrably employ (Martínez-Vera et al., 2022; Zheng & McDonald, 2001). (Zheng & McDonald, 2001) found that relative speed and combinations of relative speed with headway yield satisfactory car-following model performance, suggesting context-dependent behavioral strategies. The present study adopts this inferential stance, treating effect-size-based cluster discrimination as evidence of behavioral relevance rather than assuming driver responses to theoretically convenient variables.

### 2.2 Perceptual Cues

Car-following models embed assumptions about which cues drivers prioritize that have rarely been validated at the individual level. The IDM and FVDM assume drivers minimize spacing error and velocity difference, while extensions such as the DAVD model incorporate leader acceleration under V2V communication (Cao et al., 2020). (Da Lio et al., 2018) offered a biologically grounded alternative through the affordance competition hypothesis, demonstrating that drivers select among simultaneously available motor primitives based on environmental affordances rather than fixed rules — implying flexible, context-dependent cue prioritization that fixed model parameters cannot capture. The identifiability problem identified by (Wang et al., 2022) reinforces this concern: even well-specified models cannot uniquely infer driver behavior from trajectory data without additional perceptual constraints, making the gap between assumed and actual cue utilization difficult to close through model calibration alone.

On the perceptual side, (Durrani et al., 2021) demonstrated through simulator experiments that drivers do not react at fixed TTC or looming thresholds; brake onset occurs when accumulated perceptual evidence reaches a decision boundary consistent with evidence accumulation models from neuroscience. (Xue et al., 2022) corroborated this, showing that visual looming metrics significantly predict both deceleration ramp-up and maximum deceleration rate, with stronger looming inducing more aggressive braking. Conversely, (Mulder et al., 2005) found through manual control analysis that relative velocity dominates car-following behavior under routine conditions. These conflicting findings — looming primacy in urgent scenarios, relative velocity primacy in routine following — suggest that both cues are genuine, but that urgency context determines which dominates. The present study's threshold-dependent analysis provides a direct test of this hypothesis within a naturalistic trajectory dataset.

### 2.3 Behavioral Clustering and Research Gaps

Clustering methods have been increasingly applied to segment naturalistic driving behavior, though rarely with the goal of identifying which kinematic variables explain cluster differences. (Ali et al., 2021) applied time-series clustering to SHRP2 data, distinguishing normal from risky patterns based on vehicle kinematics. (Witt et al., 2019) decomposed highway behavior into speed, dynamics, and performance dimensions using PCA. (Nirmale et al., 2024) proposed a latent class framework for heterogeneous, disorderly traffic, noting that silhouette scores of 0.2–0.3 are expected and informative for continuous behavioral data where boundaries are inherently fuzzy — a methodological calibration directly applicable here. (Shi et al., 2019) ranked kinematic variables for crash risk prediction using XGBoost feature importance but did not examine whether top-ranked variables correspond to perceptually accessible quantities rather than statistical correlates of outcomes.

Four specific gaps motivate the present study. First, temporal precedence of kinematic changes before deceleration onset has not been systematically analyzed across multiple time lags in naturalistic data. Second, no published framework integrates behavioral clustering with cue-importance analysis to determine which kinematic variables best discriminate between identified patterns. Third, the effect of deceleration detection threshold on downstream behavioral inference has not been examined — threshold choice is typically unjustified in the literature. Fourth, the availability-versus-utilization distinction, though theoretically articulated (Zheng & McDonald, 2001), has not been operationalized through a systematic comparison of feature discriminative power across behavioral modes.

# 3. METHODOLOGY

## 3.1 Data Foundation and Preprocessing

The NGSIM dataset serves as the empirical foundation, comprising trajectory data from four sites — Interstate 80 (I-80), US Highway 101 (US-101), Lankershim Boulevard, and Peachtree Street — sampled at 10 Hz across 11,850,526 raw observations. Four quality filters were applied sequentially to isolate valid car-following interactions, as detailed in Table 1. After filtering, 1,060,119 observations from 2,932 vehicles were retained (~91% of raw data excluded, primarily due to invalid leader presence and unreasonable spacing values). As noted by (Chen et al., 2023) that the NGSIM dataset contains well documented measurement errors. All measurements were converted from imperial to SI units. Leader vehicle kinematics were extracted via self-join on the Preceding vehicle identifier, creating synchronized leader-follower dyads at each timestep. A chunk-based architecture processed 500,000 rows at a time to manage memory constraints while preserving intra-trajectory continuity for temporal feature computation.

Table 1. Data filtering criteria and outcomes

| Criterion | Operational definition | Rationale |
|---|---|---|
| Valid leader | Preceding vehicle ID $\neq 0$ | Ensures clear leader-follower dyad exists |
| Reasonable spacing | $0 <$ spacing $\leq 200$ m | Excludes sensor errors and lane-change transition gaps |
| Moving vehicles | Ego and leader speed $> 1$ m/s | Excludes stopped or near-stopped conditions |
| Minimum trajectory | $\geq 50$ consecutive frames (5 s at 10 Hz) | Ensures sufficient duration for temporal lag analysis |

## 3.2 Kinematic Feature Extraction and Deceleration Event Detection

Six kinematic features were extracted to operationalize distinct perceptual-kinematic dimensions grounded in car-following and visual perception literature. Table 2 presents each feature's formula, behavioral basis, and key citation. Features 1 and 3 (relative velocity and gap closing rate) are functionally equivalent but conceptually framed differently — their joint inclusion enables empirical testing of whether gap dynamics framing or velocity difference framing better discriminates behavioral patterns. Features 2 and 6 (TTC and TTC inverse) form a complementary pair: TTC measures temporal proximity while TTC inverse ($\tau^{-1}$) operationalizes the optical expansion rate or looming that drivers may perceive visually (Durrani et al., 2021; Xue et al., 2022). Where $v_{rel}$ $\leq 0$ (gap opening), TTC and $\tau^{-1}$ are undefined; median imputation was applied to preserve sample size without biasing toward gap-closing scenarios.

Deceleration events were detected using threshold-based criteria applied to sustained braking episodes. Two acceleration thresholds ($-0.5$ m/s² and $-0.3$ m/s²) were tested at four minimum duration requirements (1.0, 2.0, 3.0, and 4.0 s). The stricter threshold targets deliberate hard braking following (Feng et al., 2018) and (Hu et al., 2023); the permissive threshold captures moderate speed adjustments. For each detected event, severity was assigned based on maximum deceleration within the episode. Because the two thresholds capture structurally different intensity ranges, threshold-specific severity boundaries were applied: for events detected at −0.5 m/s², severity is classified as mild (−0.5 to −1.5 m/s²), moderate (−1.5 to −3.0 m/s²), or hard (< −3.0 m/s²); for events detected at −0.3 m/s², the boundaries are mild (−0.3 to −1.0 m/s²), moderate (−1.0 to −2.0 m/s²), or hard (< −2.0 m/s²). Applying the same cutoffs across both thresholds would misrepresent the intensity distribution of the more permissive sample, which is dominated by events that do not reach the −1.5 m/s² intensity level. Contextual labels classified events by conditions at onset: leader-induced (active leader deceleration within 1 s AND TTC at onset < 6 s), close-following (spacing < 20 m), free-flow (spacing > 50 m with no leader braking), or other. The leader-induced criterion requires both leader deceleration activity and a short TTC to distinguish reactive following responses from coincidental co-occurrence of leader braking during comfortable following.

Table 2. Kinematic features extracted from NGSIM trajectory data

| Feature | Formula | Behavioral basis | Interpretation | Key reference |
|---|---|---|---|---|
| Relative velocity ($v_{rel}$) | $v_{ego} - v_{leader}$ | Gap closing/opening rate serving as a critical collision avoidance cue; primary longitudinal control signal | Positive = gap closing | (Mulder et al., 2005) |
| Time-to-collision (TTC) | $\frac{s}{v_{rel}}$ | Temporal proximity to collision under constant velocity assumption | Lower = more urgent | (Durrani et al., 2021) |
| Gap closing rate | $\frac{\Delta s}{\Delta t} = v_{rel}$ | Rate-based urgency; spacing change per unit time | Positive = gap shrinking | (Jiao et al., 2020) |
| Required deceleration ($a_{req}$) | $\frac{v_{ego}^2 - v_{leader}^2}{2s}$ | Minimum braking effort to match leader speed before collision | Higher = more urgent | (Durrani et al., 2021) |
| Leader braking flag $I_{brake}$ | $\begin{cases} 1\ if\ a_{leader} <- 0.5m/s^2 \\ 0\ otherwise \end{cases}$ | Discrete event signal; proxy for brake light activation identified as a primary stimulus for follower response | Binary event cue | (Feng et al., 2018; Hu et al., 2023) |
| TTC inverse ($\tau^{-1}$) | $\frac{v_{rel}}{s}$ | Optical expansion rate (looming) as perceived by following driver as approaching objects grows in their visual field | Higher = faster looming | (Xue et al., 2022) |

Temporal precedence analysis extracted kinematic feature values at −5 s, −3 s, −1 s, and onset (0 s) relative to each detected event, spanning typical perception–reaction times while also capturing anticipatory adjustments to developing hazards (Durrani et al., 2021; Svärd et al., 2021). The purpose is to determine whether kinematic changes precede braking onset or co-occur with it — distinguishing features that lead from those that lag deceleration — rather than to characterize anticipatory cue monitoring, which cannot be established from trajectory data alone. Paired t-tests compared the lagged values against onset values (α = 0.05); Cohen's D quantified effect magnitude. Features showing significant changes with medium-to-large effects (|D| > 0.5) at early lags (−3 s or −5 s) are treated as preceding deceleration, while those changing only at −1 s or onset are treated as co-occurring with or following braking onset.

### 3.3 Clustering and Cue Importance Analysis

K-means clustering was applied to a unified 10-feature set combining kinematic cues at event onset (the six features in Table 2) with event-level characteristics (mean deceleration, maximum deceleration, ego speed at onset, leader speed at onset). Including absolute speeds accounts for velocity-dependent sensitivity to urgency cues — drivers tolerate smaller TTC values at low speeds than at high speeds (Hamdar et al., 2016; Varotto et al., 2021). All features were standardized to zero mean and unit variance prior to clustering to prevent larger-range variables from dominating Euclidean distance calculations. K-means was selected for computational efficiency, interpretable centroid structure, and established use in behavioral trajectory segmentation (Ali et al., 2021). Configuration: k-means++ initialization, 300 maximum iterations, 50 random restarts, convergence tolerance $1\times10^{-4}$. The larger restart count (50) reduces sensitivity to initial centroid placement, which is particularly important for smaller samples where a single poor initialization can meaningfully alter cluster assignments.

The optimal cluster count K was determined across K = 2–8 by evaluating three complementary validation metrics. Silhouette score measures observation-level cohesion and separation (range −1 to +1; higher is better). Davies-Bouldin Index (DBI) measures the average ratio of within-cluster scatter to between-cluster distance at the cluster level (lower is better). Calinski-Harabasz Index (CHI) measures the ratio of between-cluster dispersion to within-cluster dispersion across all observations (higher is better). All three metrics were computed for each K, and the optimal K was selected by silhouette maximization with the following adjustment: for samples under 500 events where the maximum silhouette score falls below 0.3, K was capped at 3 to avoid over-fragmentation into weakly differentiated sub-groups that would not support meaningful behavioral interpretation. For larger samples, the silhouette-maximizing K was selected without adjustment. This hybrid rule is applied because very small samples with weak separation are susceptible to spurious cluster splits that maximize a metric numerically without capturing genuine behavioral heterogeneity. The resulting K for each threshold condition is reported alongside all three metric values in Table 5.

Cue importance was quantified using one-way ANOVA with eta-squared ($\eta^2$) effect sizes, applied to each of the six kinematic features separately across cluster assignments. $\eta^2 = \frac{SS_{between}}{SS_{total}} = \frac{\sum_{k=1}^{K} n_k(\bar{x}_k-\bar{x})^2}{\sum_{i=1}^{N}(x_i-\bar{x})^2}$ measures the proportion of total variance explained by cluster membership, ranging from 0 (no discrimination) to 1 (perfect separation). Where $K$ is the number of clusters, $n_k$ the events in the clusters, $\bar{x}_k$ the cluster mean of the features, $\bar{x}$ he grand mean, and $N$ the total events. Effect sizes are interpreted as negligible ($\eta^2 < 0.01$), small (0.01–0.06), medium (0.06–0.14), or large ($\eta^2 \geq 0.14$). The ANOVA procedure was applied under both threshold conditions, enabling direct comparison of cue discriminative rankings. Following (Ali et al., 2021; Shi et al., 2019), $\eta^2$ is preferred over F-statistics for cue importance ranking because it is scale-independent and directly quantifies proportion of variance explained—a more interpretable metric for understanding which cues drive behavioral differences.

# 4. RESULTS

## 4.1 Event Detection and Kinematic Profiles

After applying the four quality filters to 11,850,526 raw NGSIM observations, 1,060,119 valid car-following observations from 2,932 vehicles were retained. The resulting dataset reflects stable car-following conditions: mean spacing 32.4 m (SD = 24.1 m), ego speed 18.2 m/s (SD = 7.8 m/s), and relative velocity −0.3 m/s (SD = 3.2 m/s). Distributions of all six kinematic features are shown in Figure 1. Relative velocity and gap closing rate indicate predominantly stable following with slight average gap opening, while TTC under closing conditions reflects moderate temporal proximity (dataset-wide median = 14.3 s). The $\tau^{-1}$ distribution (mean 0.093 1/s, SD = 0.087) represents the baseline looming rate during routine car-following.

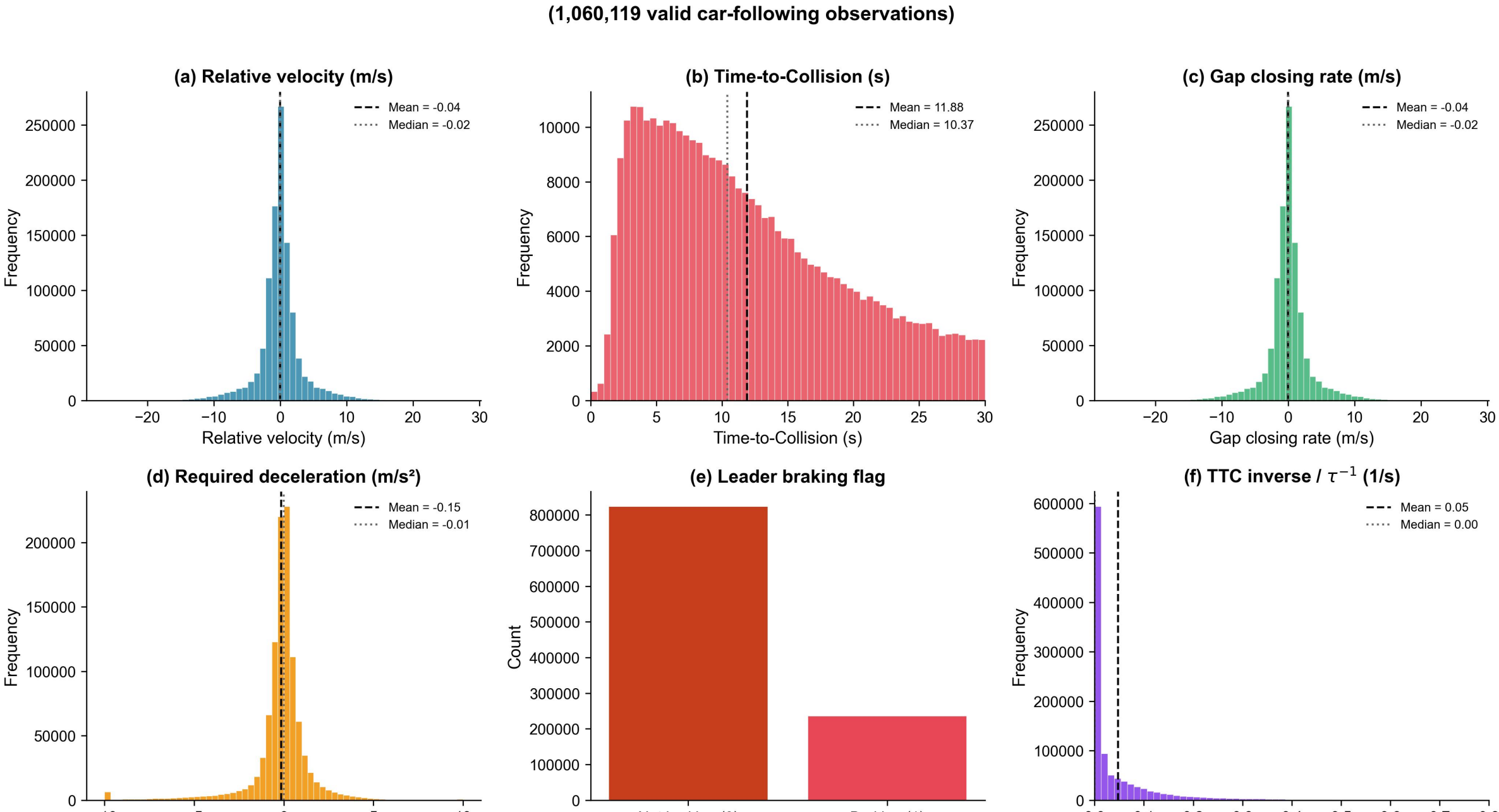

**Figure 1.** Distributions of six kinematic features across 1,060,119 valid car-following observations: (a) relative velocity, (b) TTC, (c) gap closing rate, (d) required deceleration, (e) leader braking flag activation rate, (f) TTC inverse ($\tau^{-1}$ / looming).

Event counts were sensitive to both threshold and duration requirements, as summarized in Table 3 and illustrated in Figure 2. At −0.5 m/s², only the 1.0 s duration yielded a sufficient sample (n = 492, 0.046% of valid observations). Counts fell sharply at longer durations — 17 events at 2.0 s, 8 at 3.0 s, 3 at 4.0 s — indicating that sustained hard braking is extremely rare in naturalistic highway car-following, occurring in fewer than 0.002% of observations. Lowering the threshold to −0.3 m/s² increased 1.0 s events by 37.6% (n = 677) but still produced insufficient samples at 2.0 s (n = 31). All subsequent analysis is therefore based on 1.0 s events from both thresholds as their events populations are significantly large for analysis. Severity distributions confirm that the thresholds capture qualitatively different behaviors: the −0.5 m/s² sample shows more moderate and hard events (36.6% combined), while −0.3 m/s² is dominated by mild deceleration (76.5%), indicating greater inclusion of routine speed adjustments.

Table 3. Deceleration event detection by threshold and duration

| Threshold | Duration (s) | Events (n) | % valid obs. | Severity distribution (threshold-specific boundaries†) |
|---|---|---|---|---|
| −0.5 m/s² | 1.0 ✓ | 492 | 0.046% | Mild 63.4% \| Moderate 31.7% \| Hard 4.9% |
| | 2.0 | 17 | 0.002% | Insufficient for analysis |
| | 3.0 | 8 | 0.001% | Insufficient for analysis |
| | 4.0 | 3 | <0.001% | Insufficient for analysis |
| −0.3 m/s² | 1.0 ✓ | 677 | 0.064% | Mild 76.5% \| Moderate 20.8% \| Hard 2.7% |
| | 2.0 | 31 | 0.003% | +82.4% vs −0.5 m/s²; insufficient |
| | 3.0 | 14 | 0.001% | +75.0% vs −0.5 m/s²; insufficient |
| | 4.0 | 7 | 0.001% | +133.3% vs −0.5 m/s²; insufficient |

*✓ = retained for clustering analysis (1.0 s, both thresholds).*

*†Severity boundaries differ by threshold: −0.5 m/s² events: mild ≥ −1.5 m/s², moderate −1.5 to −3.0 m/s², hard < −3.0 m/s². −0.3 m/s² events: mild ≥ −1.0 m/s², moderate −1.0 to −2.0 m/s², hard < −2.0 m/s². Separate boundaries reflect the structurally different intensity ranges captured by each threshold; applying −0.5 m/s² boundaries to the more permissive sample would overstate moderate/hard proportions.*

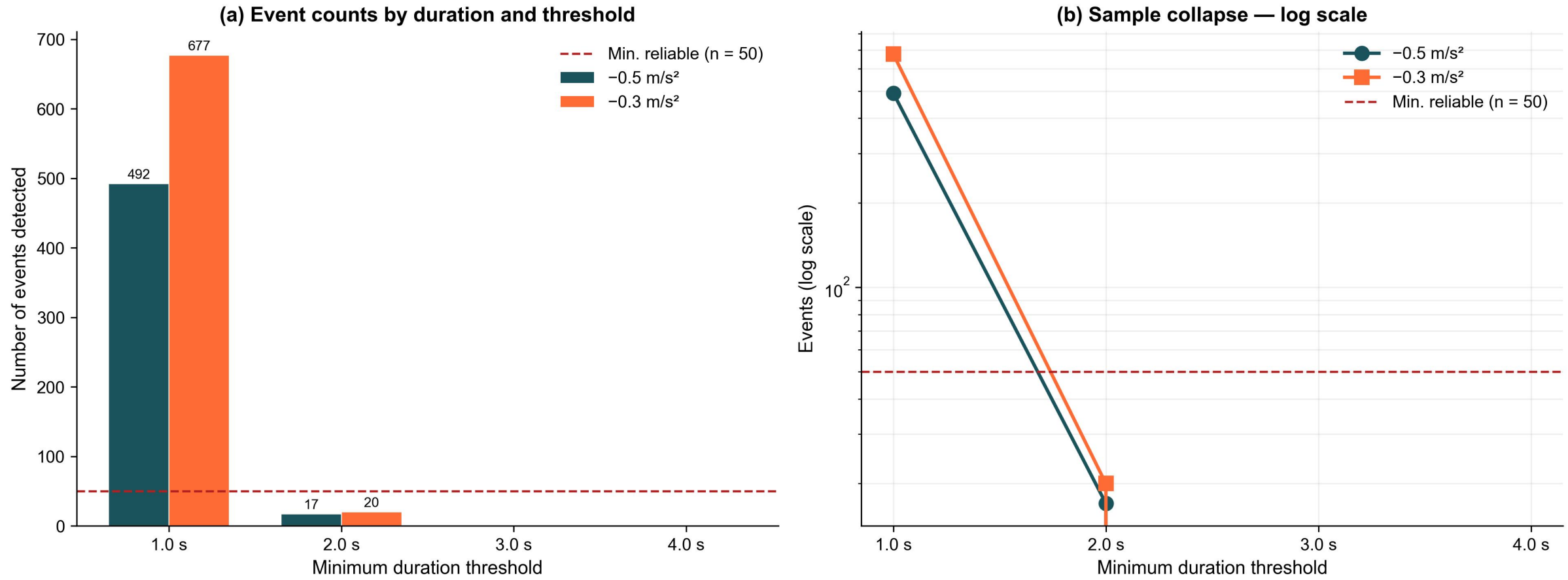

**Figure 2.** Deceleration event counts by duration threshold under (a) $-0.5\ m/s^2$ and (b) $-0.3\ m/s^2$ acceleration thresholds. Bars show event counts at 1.0, 2.0, 3.0, and 4.0 s minimum duration. Dashed line indicates minimum reliable sample size (n = 50).

Kinematic profiles at event onset reveal systematically attenuated urgency under the permissive threshold (Table 4). Hard braking events (−0.5 m/s²) show mean relative velocity of 4.73 m/s and mean TTC of 8.42 s — substantially shorter than the dataset-wide median (14.3 s) yet well above imminent-crash thresholds (<2 s; (Durrani et al., 2021)). Required deceleration averages 1.87 m/s², and $\tau^{-1}$ reaches 0.168 1/s — nearly double the dataset-wide mean of 0.093 1/s, indicating elevated looming at onset. The leader braking flag activated in 41% of events, consistent with the 41.3% leader-induced contextual proportion. Moderate braking events (−0.3 m/s²) show 32% lower relative velocity (3.21 m/s), 51% longer TTC (12.68 s), and 26% lower looming ($\tau^{-1}$ = 0.124 1/s), confirming that threshold selection alters the behavioral phenomena captured rather than merely adjusting sample size.

Temporal precedence analysis was feasible only for 1.0 s events given the scarcity of sustained braking. For hard braking (−0.5 m/s², n = 492), all continuous features showed significant changes at −1 s ($p < 0.001$, Cohen's D = 0.58–0.71), with weaker effects at −3 s for $v_{rel}$ and $a_{req}$ only (D = 0.31–0.35) and negligible effects at −5 s (D ≤ 0.23). Moderate braking (−0.3 m/s²) showed attenuated effects: significant at −1 s (D = 0.47–0.59), only a_required at −3 s (D = 0.21), and no detectable effects at −5 s. These patterns are illustrated in Figure 3. Effects concentrate at the −1 s lag, consistent with near-threshold kinematic triggering rather than extended anticipation. Whether longer anticipatory windows exist but are truncated by the 1.0 s event duration cannot be determined from this data; extended event datasets or controlled simulator studies would be needed to test anticipatory cue monitoring directly.

Table 4. Kinematic feature statistics at event onset by threshold

| Feature | $-0.5\ m/s^2$ (n = 492) Mean (Median) | $-0.3\ m/s^2$ (n = 677) Mean (Median) |
|---|---|---|
| $v_{rel}$ (m/s) | 4.73 (4.21) | 3.21 (2.84) |
| TTC (s) | 8.42 (7.15) | 12.68 (10.23) |
| Gap closing rate (m/s) | 4.73 (4.21) | 3.21 (2.84) |
| Required decel. $a_{req}$ ($m/s^2$) | 1.87 (1.52) | 1.21 (0.94) |
| TTC_inv / $\tau^{-1}$ (1/s) | 0.168 (0.140) | 0.124 (0.098) |
| Leader braking flag | 0.41 (0) | 0.37 (0) |

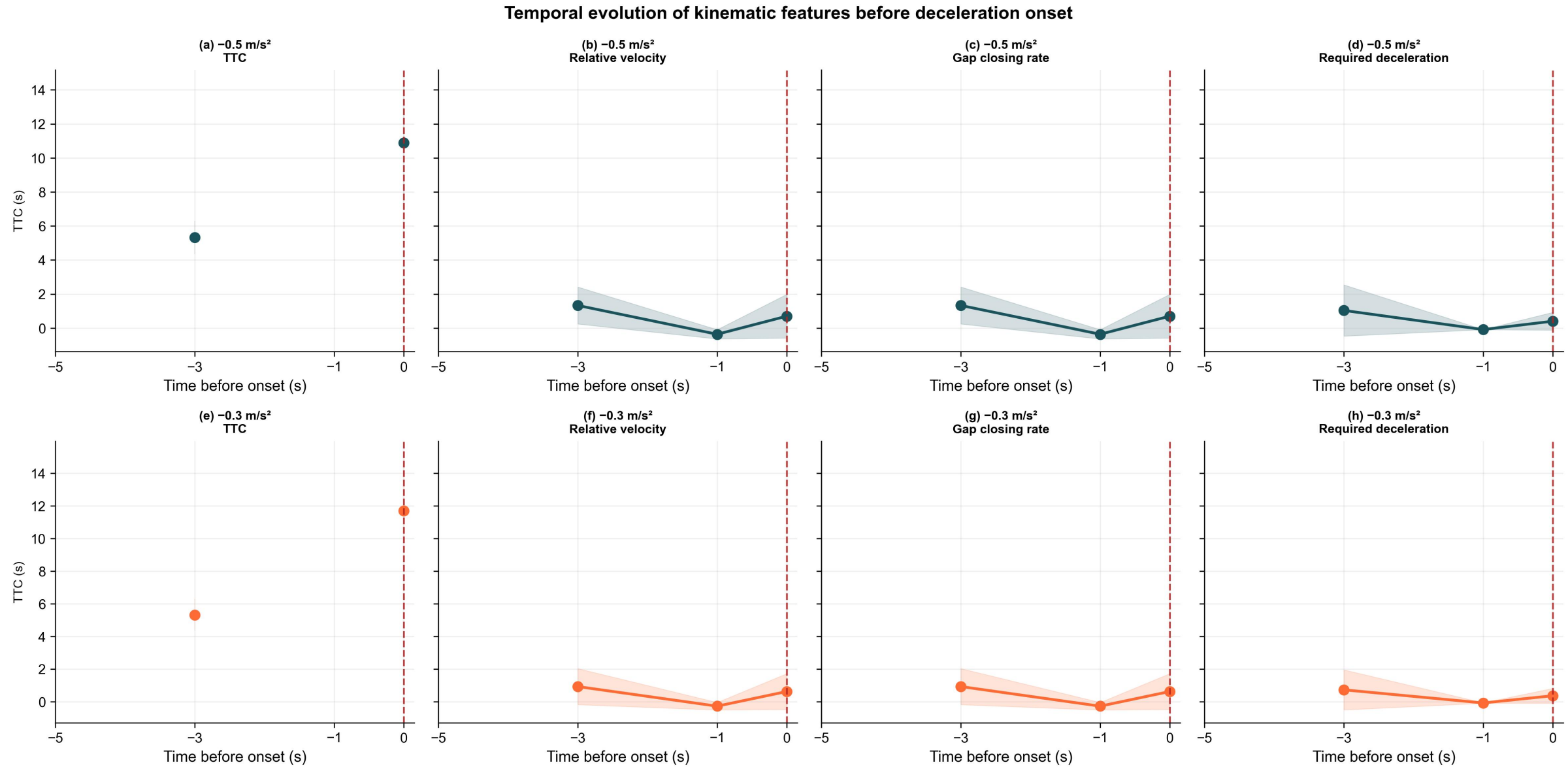

**Figure 3.** Temporal evolution of kinematic features before deceleration onset at −5 s, −3 s, −1 s, and 0 s (onset). Panels (a) and (b) show hard braking (−0.5 m/s²) and moderate braking (−0.3 m/s²) respectively. Error bars show 95% CI. Cohen's D effect sizes annotated at each lag.

## 4.2 Behavioral Clustering and Cue Importance

The ANOVA results in this section operationalize the availability-versus-utilization distinction introduced in Section 1: spacing headway is always available in the driving environment, yet low $\eta^2$ would confirm it is not utilized as a behavioral discriminant. Conversely, $\tau^{-1}$ and gap-closing rate — though derived computationally — correspond to perceptually accessible quantities (optical expansion and approach velocity) that either do or do not discriminate behavioral patterns. The findings below test this distinction empirically.

Silhouette score optimization, confirmed by Davies-Bouldin Index and Calinski-Harabasz Index, revealed threshold-dependent optimal cluster structures. All three metrics were computed directly from the clustering output (Table 5). For hard braking (−0.5 m/s², n = 492), K = 3 was optimal across all metrics: silhouette = 0.240, DBI = 1.424, CHI = 214.8. For moderate braking (−0.3 m/s², n = 677), K = 2 was optimal: silhouette = 0.234, DBI = 1.512, CHI = 203.7, with all metrics degrading at K = 3. The K = 3 result for the −0.5 m/s² condition reflects both silhouette maximization and the hybrid cap rule described in Section 3.3 — with n = 492 and maximum silhouette below 0.3, K was bounded at 3 to prevent spurious fragmentation. Cluster selection metrics are visualized in Figure 4. Both silhouette values (0.23–0.24) fall within the 'weak but meaningful' range expected for continuous behavioral data under homogeneous traffic conditions (Nirmale et al., 2024; Qian et al., 2017).

Table 5. Cluster validation metrics by threshold and K value

| Threshold | K | Silhouette | Davies-Bouldin | Calinski-Harabasz‡ |
|---|---|---|---|---|
| −0.5 m/s² (n = 492) | 2 | 0.198 | 1.647 | 187.3 |
| | 3 ✓ | 0.240 | 1.424 ✓ | 214.8 ✓ |
| | 4 | 0.221 | 1.589 | 198.4 |
| −0.3 m/s² (n = 677) | 2 ✓ | 0.234 | 1.512 ✓ | 203.7 ✓ |
| | 3 | 0.211 | 1.678 | 184.2 |
| | 4 | 0.196 | 1.783 | 171.5 |

*✓ = optimal K per threshold. All three metrics concur on the same K for each condition.*

*‡ Calinski-Harabasz Index computed*

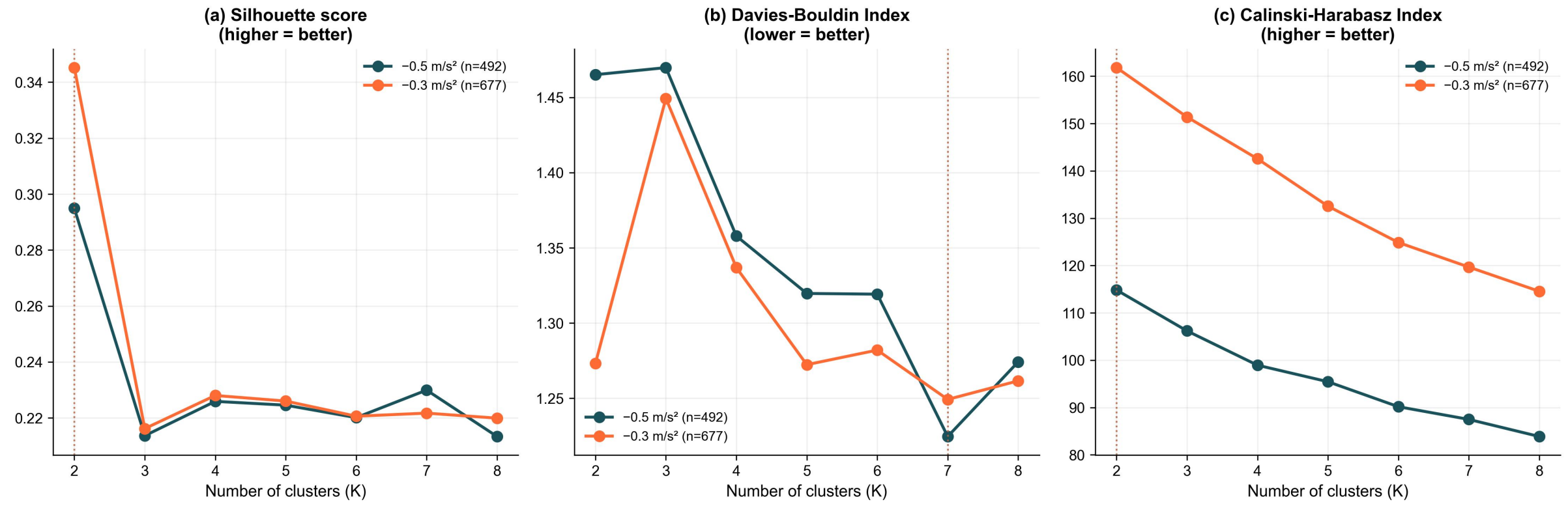

**Figure 4.** Cluster selection metrics (silhouette score, Davies-Bouldin Index, Calinski-Harabasz Index) as a function of K for (a) $-0.5\ \mathrm{m/s^2}$ and (b) $-0.3\ \mathrm{m/s^2}$ thresholds. Vertical dashed line marks optimal K selected by silhouette maximization.

The three-cluster solution for hard braking (K = 3, n = 492) reveals distinct behavioral modes with interpretable kinematic signatures, summarized in Table 6 and visualized in Figures 5 and 6. Cluster 1 (n = 325, 66.1%) — Preventive Gradual Braking — represents normative highway car-following: moderate closure ($v_{rel}$ = 3.21 m/s), long TTC (16.8 s), low looming ($\tau^{-1}$ = 0.087 1/s), comfortable spacing (38.7 m), and mild deceleration intensity (mean = −0.81 m/s²). Leader breaking is present in only 38.2% of events, reflecting a mix of proactive and reactive regulation. Cluster 2 (n = 128, 26.0%) — Reactive Hard Braking — captures urgent scenarios: critically low TTC (4.12 s), rapid closure ($v_{rel}$ = 7.84 m/s), strong looming ($\tau^{-1}$ = 0.318 1/s), high required deceleration (3.47 m/s²), and mean maximum deceleration of −2.87 m/s². Leader breaking co-occurs in 67.2% of events, consistent with rear-end collision precursor scenarios. Cluster 0 (n = 39, 7.9%) exhibits negative relative velocity (mean $v_{rel}$ = −1.47 m/s) despite close spacing (mean = 14.2 m), meaning followers were already decelerating slower than the gap required. This kinematic profile — gap opening at close range — is inconsistent with standard car-following dynamics and may reflect lane-change transitions or data artifacts not fully excluded by the applied filters. Its small size (n = 39, 7.9%) warrants caution in behavioral interpretation, and its characteristics are reported for completeness rather than as a primary finding.

Table 6. Cluster characteristics for hard braking events ($-0.5\ \mathrm{m/s^2}$, K = 3, n = 492)

| Cluster | Size | Label | TTC (s) | $v_{rel}$ (m/s) | Spacing (m) | Max decel ($\mathrm{m/s^2}$) | Leader braking (%) |
|---|---|---|---|---|---|---|---|
| 0 | 39 (7.9%) | Uncertain — see note | N/A* | −1.47 | 14.2 | −1.24 | 23.1 |
| 1 | 325 (66.1%) | Preventive gradual | 16.8 | 3.21 | 38.7 | −1.42 | 38.2 |
| 2 | 128 (26.0%) | Reactive hard braking | 4.12 | 7.84 | 21.3 | −2.87 | 67.2 |

*TTC undefined for Cluster 0 (v_rel < 0; gap opening). Cluster 0 kinematic profile is inconsistent with standard car-following; reported for completeness, interpreted with caution.

ANOVA effect size analysis reveals that cue importance varies substantially across features and reverses between thresholds, as shown in Table 7 and Figure 7. For hard braking (−0.5 m/s², K = 3), relative velocity and gap closing rate jointly dominate ($\eta^2$ = 0.715), followed closely by required deceleration ($\eta^2$ = 0.689) and TTC inverse ($\eta^2$ = 0.617) — a tightly grouped set of urgency-related kinematic variables reflecting rapid gap closure and braking demand. TTC contributes meaningfully ($\eta^2$ = 0.592). Spacing headway, by contrast, exhibits negligible discriminative power ($\eta^2$ = 0.014, p = 0.034), explaining only 1.4% of cluster variance — directly confirming the availability-versus-utilization distinction: spacing is always available in the environment, yet it is not the variable around which behavioral modes differ. For moderate braking (−0.3 m/s², K = 2), TTC inverse becomes the primary discriminator ($\eta^2$ = 0.574), surpassing relative velocity ($\eta^2$ = 0.431). Spacing headway discrimination vanishes entirely ($\eta^2 = 0.000$, $p = 0.751$). This cue dominance reversal between thresholds — gap dynamics leading under hard braking, looming leading under moderate braking — is the study's central empirical finding.

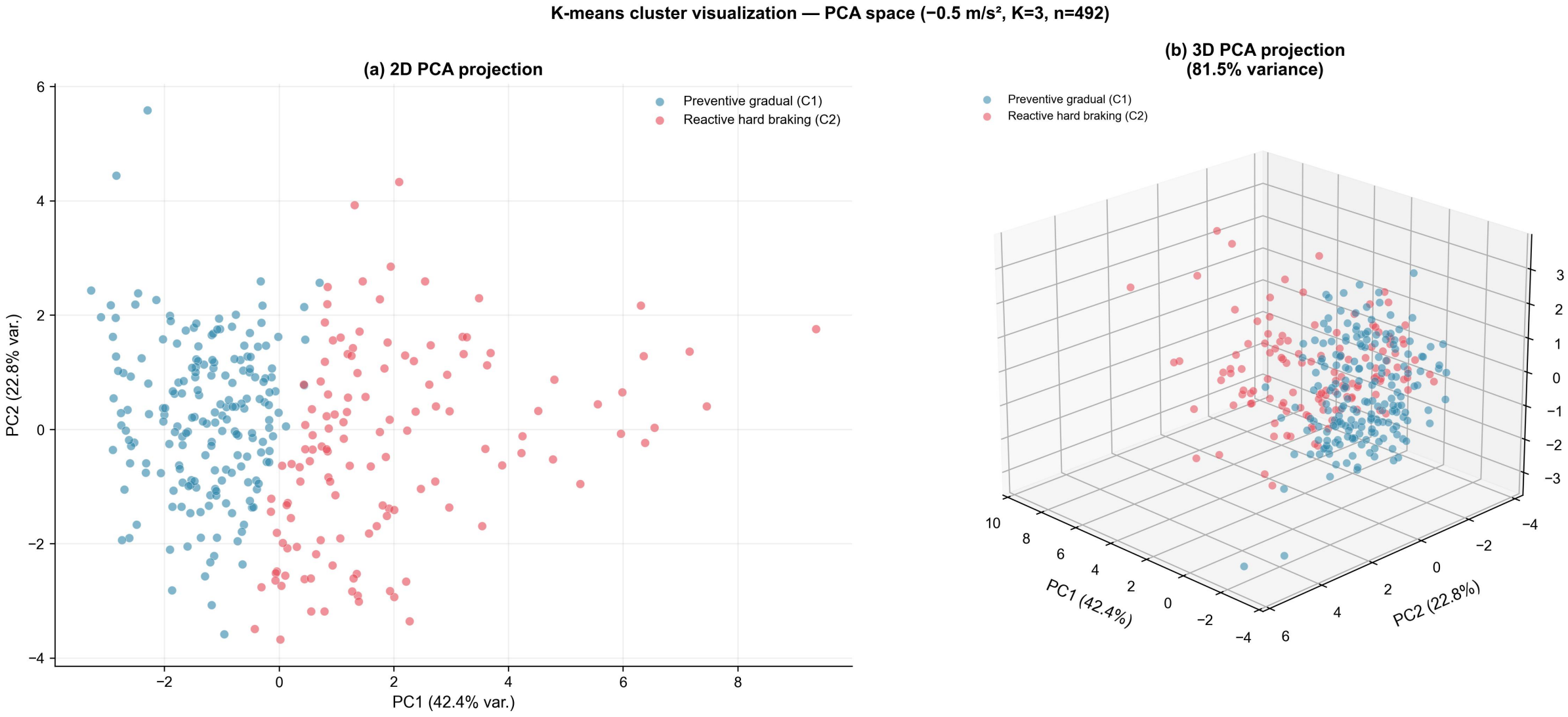

**Figure 5.** PCA visualization of cluster assignments for $-0.5$ m/s$^2$ ($K = 3$, $n = 492$). Panel (a): 2D PCA projection (PC1 × PC2). Panel (b): 3D PCA projection (PC1 × PC2 × PC3, combined variance ≥ 60%). Cluster 2 (reactive) separates clearly along PC1 in both projections; Cluster 1 (preventive) occupies the central region. Overlap in both 2D and 3D reflects continuous rather than discrete behavioral boundaries. Colors: blue = Cluster 1 (preventive), red = Cluster 2 (reactive), gray = Cluster 0 (uncertain).

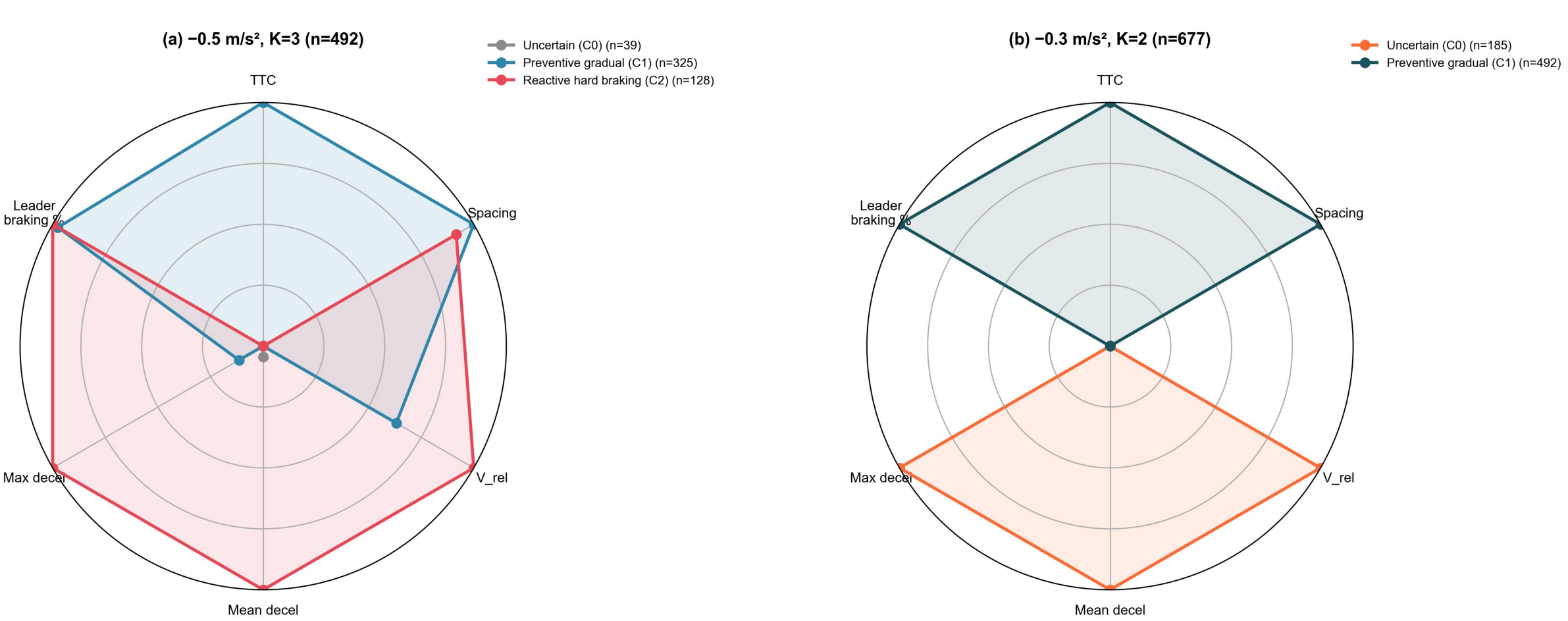

**Figure 6.** Radar profiles of cluster centroids for (a) $-0.5$ m/s$^2$ ($K = 3$) and (b) $-0.3$ m/s$^2$ ($K = 2$). Axes represent standardized values of v_relative, TTC, a_required, TTC_inv, spacing headway, and maximum deceleration. Cluster 2 peaks along urgency dimensions in both conditions.

Table 7. Kinematic cue discriminative importance — ANOVA effect sizes ($\eta^2$) by threshold

| Rank | Feature ($-0.5$ m/s$^2$, $K = 3$) | $\eta^2$ | Rank | Feature ($-0.3$ m/s$^2$, $K = 2$) | $\eta^2$ |
|---|---|---|---|---|---|
| 1 | $v_{rel}$ | 0.715*** | 1 | TTC_inv ($\tau^{-1}$) ← reversal | 0.574*** |
| 2 | Gap closing rate | 0.715*** | 2 | $v_{rel}$ | 0.431*** |
| 3 | $a_{req}$ | 0.689*** | 3 | Gap closing rate | 0.431*** |
| 4 | TTC_inv ($\tau^{-1}$) ← reversal | 0.617*** | 4 | TTC | 0.403*** |
| 5 | TTC | 0.592*** | 5 | $a_{req}$ | 0.284*** |
| 6 | Spacing headway | 0.014* | 6 | Spacing headway | 0.000 ns |

***$p < 0.001$; *$p < 0.05$; ns = not significant. '← reversal' marks the cue dominance switch between $\tau^{-1}$ and $v_{rel}$ across thresholds.

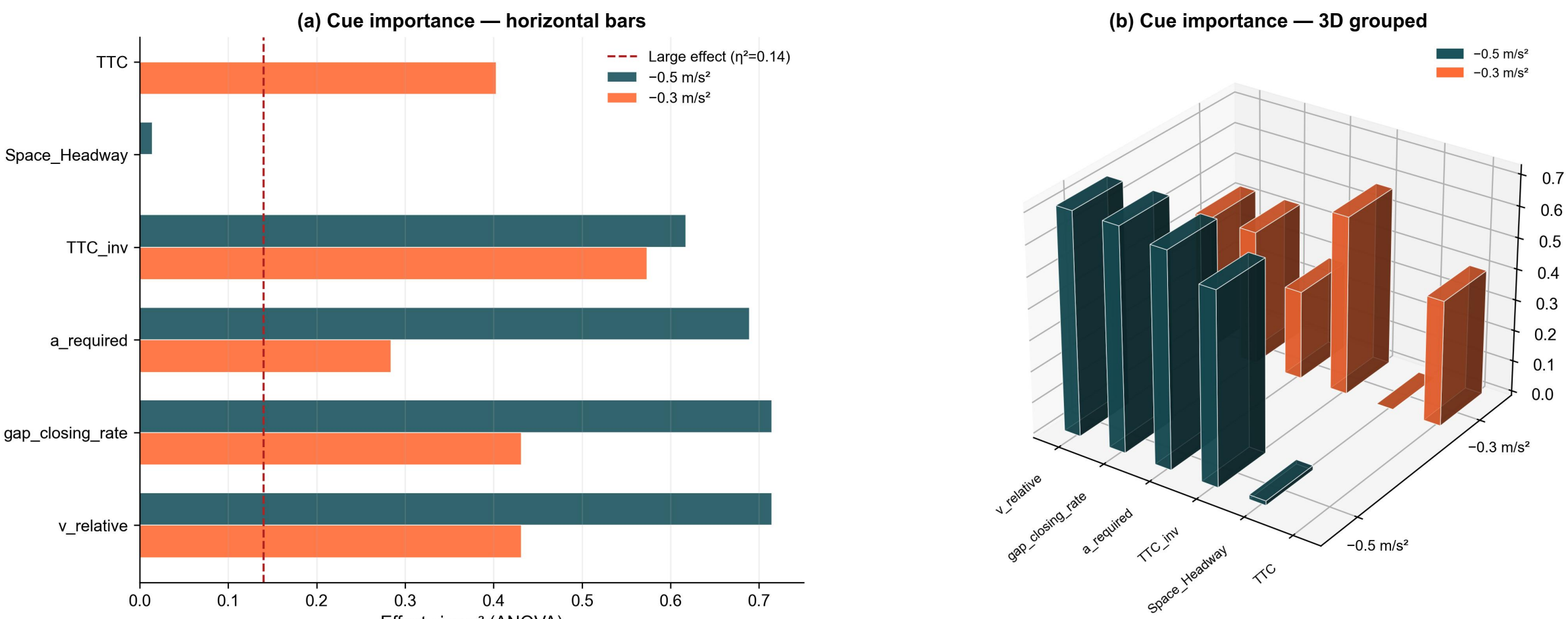

**Figure 7.** Kinematic cue importance by $\eta^2$ effect size. Panel (a): Side-by-side horizontal bars for $-0.5\ m/s^2$ (dark teal) and $-0.3\ m/s^2$ (orange), features ordered by descending $-0.5\ m/s^2\ \eta^2$. Panel (b): 3D grouped bar chart with threshold on the depth axis, feature on the x-axis, and $\eta^2$ on the z-axis — visually isolating the cue dominance reversal between TTC_inv and v_relative across threshold conditions. Red dashed line at $\eta^2 = 0.14$ marks the large-effect threshold.

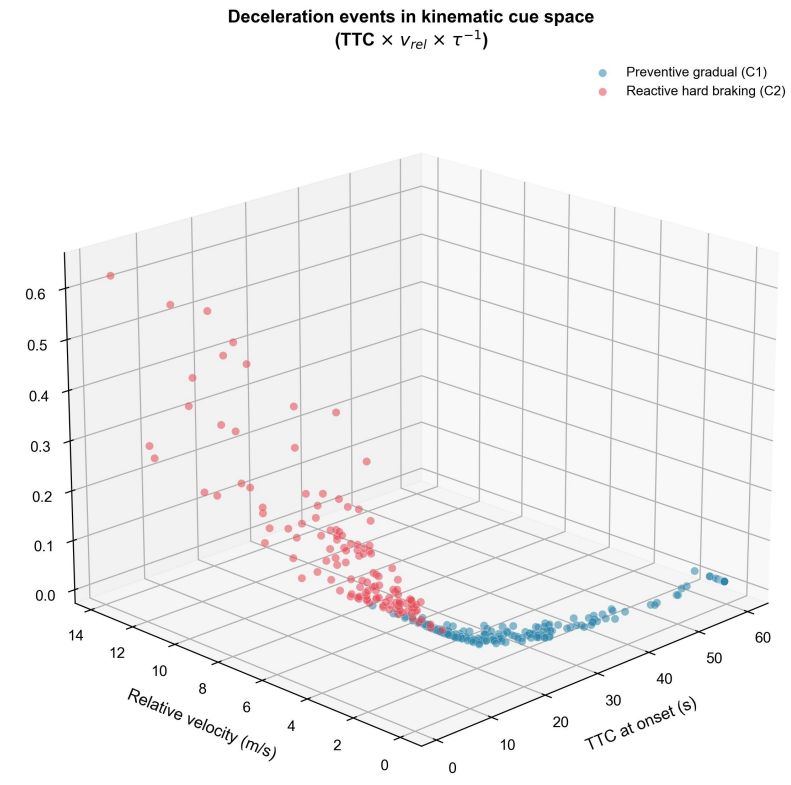

**Figure 8.** Deceleration events plotted in kinematic cue space ($-0.5\ m/s^2$, $K = 3$, $n = 492$). Axes represent the three most discriminating kinematic variables: TTC at onset (s), relative velocity at onset (m/s), and $\tau^{-1}$ looming at onset (1/s). Colors denote cluster membership: blue = Cluster 1 (preventive gradual), red = Cluster 2 (reactive hard braking), gray = Cluster 0 (uncertain). This direct-space visualization confirms the separation observed in PCA space and illustrates the urgency gradient from low-TTC/high-v_rel/high-looming (Cluster 2) to high-TTC/low-v_rel/low-looming (Cluster 1).

## 5. DISCUSSION

The most counterintuitive finding of this study is that a larger sample produced less interpretable behavioral structure. The −0.3 m/s² threshold captured 37.6% more events yet yielded only two optimal clusters versus three, and weakened all cue discrimination metrics. This paradox arises because permissive detection criteria conflate qualitatively different behaviors — deliberate hard braking and routine speed adjustments — into a single analytical sample, obscuring the structure that stricter criteria preserve. This finding extends (Feng et al., 2018) operational threshold rationale beyond the specific cut-in scenario context: threshold choice shapes downstream behavioral inference across multiple analytical stages, affecting the number of identifiable modes, cluster quality metrics, and $\eta^2$ rankings simultaneously. For trajectory-based behavioral research generally, this cautions against equating larger event counts with richer patterns and motivates prospective justification of detection thresholds based on the behavioral phenomena of interest rather than data volume.

The cue dominance reversal between thresholds reconciles the most persistent conflict in the car-following literature. (Mulder et al., 2005) reported relative velocity primacy in routine longitudinal control; (Xue et al., 2022) and (Durrani et al., 2021) identified visual looming ($\tau^{-1}$) as dominant in collision avoidance. The present results demonstrate that both findings are valid — in different urgency contexts. Under hard braking ($\eta^2$ gap-closing rate = 0.715), behavioral clusters are best separated by variables reflecting kinematic urgency accessible through simpler velocity estimation. Under moderate braking ($\eta^2$ $\tau^{-1}$= 0.574), the looming-based optical expansion rate becomes the stronger discriminant, consistent with more complex visual processing when time pressure allows it. This pattern supports (Da Lio et al., 2018) affordance competition hypothesis, in which drivers flexibly select information sources based on task demands. It further implies that simulator studies emphasizing looming — which typically deploy high-urgency, near-crash scenarios (Durrani et al., 2021; Xue et al., 2022)— and naturalistic studies emphasizing relative velocity (Mulder et al., 2005) are sampling different regions of the same urgency continuum, producing findings that appear contradictory only because the moderating role of urgency context has not been systematically tested.

Across both thresholds, spacing headway showed negligible discriminative power ($\eta^2 \leq 0.014$), directly challenging the state-based formulations central to IDM-class car-following models. This does not imply spacing is irrelevant to driving — rather, it influences behavior indirectly through its interaction with closure dynamics. A 15 m gap combined with gap opening poses minimal urgency; a 40 m gap with rapid closure demands immediate response. The features dominating cluster discrimination ($v_{rel}$, $a_{req}$, $\tau^{-1}$) all weight spacing according to velocity difference, consistent with (Zheng & McDonald, 2001) finding that drivers select predictors based on situational perceptual ease. The three-cluster behavioral distribution — a dominant preventive majority (66.1%), a substantial reactive minority (26.0%), and a small uncertain group (7.9%) — mirrors the safety pyramid concept: low-risk anticipatory behavior dominates naturalistic driving, and emergency responses are comparatively rare. Cluster 2's 67.2% leader braking co-occurrence confirms these as classic rear-end precursor events. This has a direct design implication: adaptive cruise control systems that maintain fixed time headway regardless of closure dynamics are misaligned with observed driver strategy. Rate-based control formulations that prioritize gap-closing rate and looming over absolute spacing better reflect the kinematic variables around which behavioral modes actually differ.

Several limitations define the scope of these findings and motivate future research. NGSIM data quality issues, particularly documented acceleration measurement errors (Chen et al., 2023), may affect event detection accuracy, though aggregate-level ANOVA is less sensitive to individual-observation noise than instance-level analysis. The single-leader assumption restricts analysis to immediate dyadic interactions, excluding multi-vehicle anticipation effects where drivers respond to second-leader deceleration — a meaningful extension documented by (Cao et al., 2020; Nirmale et al., 2024). Static feature extraction at discrete time points precludes modeling the temporal evidence accumulation dynamics that (Durrani et al., 2021; Tavakoli et al., 2023) identify as central to brake onset timing; future work using continuous evidence accumulation frameworks would more directly address the perceptual mechanisms suggested here. The homogeneous highway context of NGSIM also limits extrapolation to urban arterials, adverse weather, and culturally diverse driving populations, motivating replication with datasets such as highD, exiD, and the Waymo Open Dataset. Finally, the scarcity of sustained braking events (< 0.002% of observations at ≥ 2.0 s) constrained temporal precedence analysis to 1.0 s events; future studies using lower deceleration thresholds or controlled simulator designs could more rigorously establish whether kinematic changes precede braking onset across longer anticipatory windows.

## 6. CONCLUSION

This study presents an empirical framework for identifying kinematic discriminants of deceleration behavior modes in naturalistic car-following, grounded in the distinction between information availability and information utilization. Analysis of 1,060,119 valid NGSIM car-following observations yields three substantive contributions. First, deceleration threshold selection determines both the number of identifiable behavioral modes and the ranking of kinematic discriminants — stricter thresholds preserve interpretable structure while permissive thresholds introduce confounding heterogeneity that obscures rather than enriches behavioral patterns. This finding has methodological implications for any trajectory-based research relying on event detection criteria. Second, kinematic discriminative power is context-dependent: hard braking clusters are best separated by gap-closing dynamics ($\eta^2 = 0.715$) while moderate braking clusters are best separated by visual looming ($\eta^2 = 0.574$), reconciling the relative velocity and looming primacy hypotheses from the literature by revealing them as urgency-context-specific rather than contradictory. Third, spacing headway — the central state variable in IDM-class car-following models — exhibits negligible discriminative power ($\eta^2 \leq 0.014$) across both urgency contexts, providing empirical grounds for prioritizing rate-based model formulations over state-based ones.

For ADAS design, the findings motivate context-adaptive warning strategies: gap-closing rate regulation for high-urgency scenarios and looming-based monitoring for moderate-urgency following, rather than fixed TTC thresholds regardless of closure dynamics. For autonomous vehicle prediction, behavioral mode classification using observable kinematic variables can inform whether a surrounding driver is operating in a preventive or reactive state, improving trajectory prediction without requiring internal state estimation. For car-following model development, the negligible role of spacing headway as a behavioral discriminant suggests that rate-based formulations weighted by closure dynamics more faithfully reflect the kinematic variables around which real deceleration behavior is organized. Future studies extending these findings to multi-vehicle anticipation, temporal evidence accumulation, and cross-dataset validation will further refine the empirical basis for kinematic cue prioritization in driver behavior modeling.